\documentclass[aps,prl,10pt,twocolumn,superscriptaddress,showpacs]{revtex4}
\usepackage[bbgreekl]{mathbbol}
\usepackage{amsmath}
\usepackage{amssymb}
\usepackage{bbm}
\usepackage{graphicx}
\usepackage{framed}
\usepackage{color}
\usepackage{hyperref}
\usepackage{epstopdf}
\usepackage{dsfont}
\usepackage{amssymb}
\usepackage{amsmath}
\usepackage{amsthm}
\usepackage{wrapfig}
\usepackage{relsize}
\usepackage{bm}
\usepackage[latin1]{inputenc}
\usepackage{enumitem}
\usepackage{natbib}
\usepackage{graphicx}
\usepackage{physics}
\usepackage{hyperref}
\usepackage{xcolor}
\usepackage{subfigure}

\begin{document}

	\title{Quantum flags, and new bounds on the quantum capacity of the depolarizing channel}
	
	\author{Marco Fanizza}
	\email{marco.fanizza@sns.it}
	\author{Farzad Kianvash}
	\email{farzad.kianvash@sns.it}
	\author{Vittorio Giovannetti}	\affiliation{NEST, Scuola Normale Superiore and Istituto Nanoscienze-CNR, I-56126 Pisa, Italy}
	\date{\today}

	\begin{abstract}
	A new bound for the quantum capacity of the $d$-dimensional depolarizing channels is presented.
	Our derivation makes use of a flagged extension of the map where the receiver obtains a copy of a state $\sigma_0$ whenever the messages are transmitted without errors, and a copy of a state $\sigma_1$ when
instead the original state gets fully depolarized. By varying the overlap 
between the flags states, the resulting transformation nicely interpolates between the depolarizing map
(when $\sigma_0=\sigma_1$), and the $d$-dimensional erasure channel (when $\sigma_0$ and $\sigma_1$ have orthogonal support). 
In our analysis we compute the product-state classical capacity, the entanglement assisted capacity and,
under degradability conditions, the quantum capacity of the flagged channel. 
From this last result we get the upper bound for the depolarizing channel, which 
by a direct comparison  appears to be tighter than previous available results for $d>2$, and for $d=2$ it is tighter in an intermediate regime of noise. In particular, in the limit of large $d$
values, our findings presents a previously unnoticed $\mathcal O(1)$ correction.	\end{abstract}
	
	\maketitle

	Quantum Shannon theory \cite{HOL BOOK,WILDE BOOK} provides a characterization of the maximum transmission rates (capacities) achievable in sending classical or quantum data through a quantum channel. The depolarizing channel (DC)~\cite{DEPCH} is the simplest and most symmetric non-unitary quantum channel but still, despite the considerable efforts that have been spent on such issue~\cite{DIVI,SMSM1,SS UPP B DEP,SMI UPP B DEP,YIN UPP B DEP,FERN1,ADAMI,CERF1,RAIN1,RAIN2,SUTT UPP B DEP,lownoiseQ,distdepo}
its so-called quantum capacity \cite{L,S,Q CAP DEV} is not known. 
 DCs have a peculiar position in the theory which make them an important error model for finite dimensional systems, like qubits in a quantum computer. Indeed  by pre- and post-processing and classical communication via twirling~\cite{Twirling Hs}, 
 any other channel can be mapped into a  DC whose quantum capacity is lower than or equal to the quantum capacity of the original channel~\cite{BKN}.
 Accordingly the value of the quantum capacity of DCs can be used to bound the 
 minimum number of physical qubits needed to preserve quantum information in quantum processors and memories.
 	In the view of these facts it is clear that  the DC quantum capacity problem  is of primary importance in quantum information theory: solving it would likely help in understanding the peculiar difficulties of quantum communication and error correction. 
	
	The evaluation of most capacities cannot be performed algorithmically, since it requires in principle an infinite sequence of optimizations, at variance with the classical case \cite{noisyshannon}. For a particular kind of channels, the degradable channels, the quantum capacity is given by the one-shot quantum capacity, which is a single-letter formula \cite{degradability}. However, the DC is not degradable
	and the one-shot quantum capacity is known to be just a lower bound. 
The main result of this paper is a new analytic upper bound to the quantum capacity of the DC valid for any finite dimension, which outperforms previous results in many different regimes.
To achieve this goal we rely on  flagged extension  of quantum channels, a construction
which, in other contexts, proved to be a powerful tool, see e.g. the result on the superadditivity of coherent information reported in
Ref.~\cite{dephrasure}. In our case we define the flagged depolaring channel (FDC)  assuming that 
 if Alice sends the density matrix $\rho$, with probability $p$ Bob receives such state together with an ancillary system 
 prepared into the state $\sigma_0$, and with probability $1-p$ the completely mixed state together with the ancillary system in $\sigma_1$.
 The density matrices  
 $\sigma_0$ and $\sigma_1$ behave as flags that encode information about what happened to the input and, at variance with
previous approaches~\cite{SS UPP B DEP,YIN UPP B DEP, distdepo}, are not assumed to be necessarily orthogonal --
 when this happens Bob can know exactly if he received the original message or an error, and our FDC is equivalent to the erasure channel~\cite{Q CAP ERA}.
 By tracing out the flags, Bob effectively receives the output of a DC. This means that  FDC is  a better communication line than its associated
DC, therefore every capacity of  the former  is larger than or equal to the corresponding value of the latter. Most importantly it is possible to find $p, \sigma_0, \sigma_1$ in such a way that the FDC is degradable obtaining a bound for the quantum capacity of the 
associated DC.
When compared with previous results our findings 
provide a better estimate of the quantum capacity of the DC for all choices of $d$ and $p$, except for $d=2$. In this case the bounds in \cite{SUTT UPP B DEP, lownoiseQ} perform better at low noise, while for higher noise the new bound is better, surpassing also the one in \cite{distdepo} in an intermediate region.
Most notably the improvement increases in the large $d$ limit: the gap between the best upper bound and lower bound of the quantum capacity is given by a $\mathcal O(1)$ function of $p$ which is differentiable in $p=0$, in contrast with previous bounds for which the $\mathcal O(1)$ term of the gap is the binary entropy~$h(p)$.

\textit{Preliminaries}.-- Given a finite dimensional Hilbert space $\mathcal{H}$, we write the space of linear operators on $\mathcal{H}$ as $\mathcal{L(H)}$ and the set of density operators as $\mathfrak{S}{({\cal H})}$. The action of 
 a quantum channel $\Lambda: \mathcal{L(H}_A)\rightarrow \mathcal{L(H}_{B})$ connecting two systems described by the Hilbert spaces $\mathcal{H}_A$ and $\mathcal{H}_B$, 
    is a Completely Positive Trace Preserving (CPTP) map~\cite{HOL BOOK}
    on $\mathcal{L(H}_A)$ which  can always cast in  the Stinespring representation form,
	\begin{equation}\label{steinspring}
		\Lambda(\theta)=\tr_{E'}(U_{AE}\: \theta_A\otimes\ket{e}\bra{e}_E\: U^{\dagger}_{AE})\: ,
	\end{equation}
	where $\ket{e}_E$ is the state of environment interacting with the system $A$, and $U_{AE}$ is an
	unitary interaction acting on 
	$\mathcal{H}_A\otimes\mathcal{H}_E \cong\mathcal{H}_B\otimes\mathcal{H}_{E'}$. 
	In this setting the complementary channel $\tilde{\Lambda}: \mathcal{L(H}_A)\rightarrow \mathcal{L(H}_{E'}) $ is defined as 	the CPTP mapping \begin{equation}\label{def compl}
	\tilde{\Lambda}(\theta):=\tr_{B}(U_{AE}\: \theta_A\otimes\ket{e}\bra{e}_E\: U^{\dagger}_{AE})\: .
	\end{equation}
	The channel $\Lambda$ is said to be degradable if there exists a third CPTP  channel $W: \mathcal{L(H}_B)\rightarrow \mathcal{L(H}_{E'})$ (dubbed degrading channel)  such that $W \circ \Lambda=\tilde{\Lambda}\:.$ Similarly, it is said to be anti-degradable if instead there exists a CPTP  channel $V: \mathcal{L(H}_{E'})\rightarrow \mathcal{L(H}_{B})$ such that  $V \circ \tilde{\Lambda}={\Lambda}\: .$
	Finally we call $N$ a degradable extension of $\Lambda$ if $N$ is degradable and there is a second channel $R$ such that $R\circ N=\Lambda$.
	
The classical capacity $C(\Lambda)$ of  $\Lambda$ is the highest achievable rate at which classical data can be faithfully transmitted through such channel. Following~\cite{C CAP H,C CAP S} it can be computed as $C(\Lambda)=\lim_{n\rightarrow \infty}C_n(\Lambda)=\lim_{n\rightarrow \infty}\tfrac{1}{n}\chi(\Lambda^{\otimes n})$, with
$\chi(\Lambda)=\max_{\{p_i;\rho_i\}} \chi(\{p_i; \Lambda(\rho_i)\})$
where the Holevo quantity of an ensemble is defined as $\chi(\{p_i; \rho_i\}):= S(\sum_i p_i \rho_i)-\sum_ip_i S(\rho_i)$, $S(\rho)$ being the Von Neumann entropy.
Similarly the entanglement assisted classical capacity $C_E(\Lambda)$ measures the highest rate at which the classical information can be transmitted through $\Lambda$ when Alice and Bob share unlimited resource of entanglement. From~\cite{ENT ASIS CAP} it follows that $ C_E(\Lambda)=\max_\rho I(\rho,\Lambda),$
	where the mutual information is defined as $I(\rho,\Lambda):=S(\rho)+S(\Lambda(\rho))-S(\tilde\Lambda(\rho))\:$.
Finally the quantum capacity $Q(\Lambda)$ gives the highest rate at which quantum information can be transmitted over many uses of $\Lambda$. In this case from~\cite{Q CAP DEV,Q CAP BAR} we get
		$Q(\Lambda)=\lim_{n\rightarrow\infty}Q_n(\Lambda)=\lim_{n\rightarrow\infty}\max_{\rho \in \mathfrak{S}{({\cal H}}_A^{\otimes n})}\;\tfrac{1}{n} J(\rho,\Lambda^{\otimes n})$,
 with $
J(\rho,\Lambda^{\otimes n}):=S(\Lambda^{\otimes n}(\rho))-S(\tilde{\Lambda}^{\otimes n}(\rho))$.	
	  For a degradable channel the regularization limit on $n$ is not needed and the expression for 	 $Q(\Lambda)$  reduces to single-letter formula
		\begin{equation}\label{def cohinfo}
	Q(\Lambda)=Q_1(\Lambda):=\max_{\rho \in \mathfrak{S}{({\cal H}}_A)}\:
	J(\rho,\Lambda)\;.
	\end{equation}

\textit{The FDC model}.--
 In a standard approach to  quantum communication
  the interaction between the quantum carriers of the information and their environment, the associated  interaction time, as well as the state of environment are assumed to be known. However, it is possible to think about  scenarios where the state of environment is changing in time and it can be monitored with quantum measurements. In this setting, suppose that with probability $p_i$ the state of environment is the state $\sigma_i$, and that when this happens information carrier gets transformed by a
 a given CPT transformation $\Lambda_i$. If there was no other information except the probability distribution of environment, the complete channel would be just the weighted sum of each individual map, i.e. $\Lambda:= \sum_{i} p_i \Lambda_i$. Instead, we assume that 
 in our case
 Bob collects a copy of the environment: in this case 
  the complete channel can be written as 
	\begin{equation}\label{flag general}
		\mathbb{\Lambda} [\cdots] 
		 :=\sum_{i} p_i \Lambda_i[\cdots] \otimes 
		\sigma_i \: ,
	\end{equation}
	where now the $\sigma_i$s live on an ancillary space ${\cal H}_1$ on which
	Bob has complete access. 
More abstractly, this model can be also seen as a quantum channel with {\it quantum flags}, where with probability $p_i$ the channel acts as $\Lambda_i$ and Bob receives a quantum flag $\sigma_i$ which encodes in a quantum state the information about which channel is acting.
As $\Lambda$ can be obtained from $\mathbb{\Lambda}$ by simply tracing away the
	flags, it turns out that the capacities of the latter provide natural upper bounds for the corresponding ones 
	of the former, i.e.	
\begin{eqnarray}\label{NICEBOUND} 
Q(\Lambda) \leq Q(\mathbb{\Lambda})\;, 
	\end{eqnarray} 
	where we specified this property in the case of the quantum capacity. 
	A special example of a channel of the form~(\ref{flag general}) was considered in \cite{SS UPP B DEP, YIN UPP B DEP} where the $\sigma_i$ were assumed to be orthogonal pure states. 
Here, on the contrary we allow the $\sigma_i$'s to be mixed and not necessarily orthogonal  and focus on the case where the resulting  mapping
has the form 
	\begin{equation}\label{the channel}
		\mathbb{\Lambda}_{p}^d[\cdots]=(1-p)[\cdots]\otimes\sigma_0+p\Tr[\cdots]\tfrac{I^d}{d}\otimes\sigma_1\: .
	\end{equation}
	This channel acts on a $d$ dimensional Hilbert space and it can be expressed as in~(\ref{flag general}) with two components, the first 
	  associated with 
	the identity channel and the second  
	 associated with a completely depolarizing transformation
	that replaces every input with  the completely mixed state ${I^d}/{d}$.
	Notice however that Eq.~(\ref{the channel}) describes  a proper CPTP mapping also for
	values of $p$ larger than 1 -- indeed its Choi state~\cite{HOL BOOK,WILDE BOOK} 
		can be easily shown to be positive for any $p>0$ such that $\tfrac{p}{d^2}\sigma_0+(1-p)\sigma_1\geq 0$.
Most importantly,  irrespectively from the value of $\sigma_0$ and $\sigma_1$, by removing the flag states
from~(\ref{the channel}) via partial trace reduces to a standard DC,
	\begin{eqnarray} 
	\Lambda_p^d[\cdots] := (1-p)[\cdots]+p\Tr[\cdots]\tfrac{I^d}{d}\;. \label{dep}
	\end{eqnarray} 
Therefore, invoking the monotonicity (\ref{NICEBOUND})
	we can  upper bound the rather elusive quantum capacity of $\Lambda_p^d$, with the quantum capacity of  $\mathbb{\Lambda}_{p}^d$
	which, as we shall see in the following section, that it is relatively easy to characterize.
	
\textit{FDC capacities}.--A fundamental ingredient in studying the capacities of $\mathbb{\Lambda}_{p}^d$ is that such channel is covariant under the action of arbitrary unitary transformations $U$ of $SU(d)$, i.e.$\mathbb{\Lambda}_{p}^d[U\cdots U^\dagger]=(U \otimes I ) \mathbb{\Lambda}_{p}^d[\cdots](U^\dagger \otimes I)$, the operators $I$ being the identity on the flags. This implies that
the output von Neumann entropy associated with a generic pure input state is a constant quantity $t(p,d,\sigma_0,\sigma_1)$
which does not explicitly depend upon the specific value of $|\psi\rangle$, but only upon
the parameters that characterize the map i.e. $S(\mathbb{\Lambda}_{p}^d[\ket{\psi}\bra{\psi}])=t(p,d,\sigma_0,\sigma_1)$.
In the Supplemental Material (SM), using  the concavity properties  of $\chi(\{p_i,\rho_i\}$ and 
 $I(\rho,\Lambda)$~\cite{HOL BOOK}, the product state classical capacity of the channel  and 
  the entanglement assisted capacity  are shown to correspond to 
	\begin{eqnarray}
	C_1(\mathbb{\Lambda}_{p}^d)&=&\log d + S((1-p)\sigma_0+p\sigma_1)-t(p,d,\sigma_0,\sigma_1), \nonumber \\
		C_E(\mathbb{\Lambda}_{p}^d)&=&2\log{d}+S\big((1-p)\sigma_0+p\sigma_1\big) 		\nonumber \\
		 &-&t(p,d^2,\sigma_0,\sigma_1).
\label{NEW} 
	\end{eqnarray}
For finding the quantum capacity we restrict the problem to the case where $\sigma_1=\ket{e_1}\bra{e_1}$ is a pure state, and $\sigma_0$ is diagonalizable in that basis, i.e. $\sigma_0=c^2\ket{e_1}\bra{e_1}+(1-c^2)\ket{e^\perp_1}\bra{e^\perp_1}$.
	 For this case both $\mathbb{\Lambda}_{p}^d$ 
	 and its complementary counterpart can be parametrised by the fidelity between $\sigma_0$ and $\sigma_1$, i.e. via the parameter $c$
	 (in particular we can write  $\mathbb{\Lambda}_{p,c}^d(\rho):=(1-p)\rho\otimes \big(c^2\ket{e_1}\bra{e_1}+(1-c^2)\ket{e^\perp_1}\bra{e^\perp_1}\big)+p\tfrac{I^d}{d}\otimes\ket{e_1}\bra{e_1}$). In the SM, using a simple measurement and action channel as a candidate for the degrading channel, we showed that $\mathbb{\Lambda}_{p,c}^d$ is degradable for $c$ fulfilling the inequality	 	\begin{equation}\label{deg regime1}
	 c \leq c(p):=\sqrt{({1-2p})/({2-2p})} \: .
	 \end{equation}
	 In this regime, the quantum capacity of $\mathbb{\Lambda}_{p,c}^d$ is equal to the product state quantum capacity $Q_1$ i.e. $Q(\mathbb{\Lambda}_{p,c}^d)$=$Q_1(\mathbb{\Lambda}_{p,c}^d)$,  and
	  we should maximize the coherent information $J$ to compute $Q_1(\mathbb{\Lambda}_{p,c}^d)$. In general, maximizing the coherent information $J$ is not an easy task: in our case however 
	the problem however gets simplified  again thanks to the degradability condition of the channel. When this property holds, in fact 
	 $J(\rho,\mathbb{\Lambda}_{p,c}^d)$ is concave in the input state $\rho$~\cite{YARD CONC}, which by covariance of $\mathbb{\Lambda}_{p,c}^d$ implies that the coherent information is maximized on the maximally mixed state
	\begin{eqnarray}
		&Q(\mathbb{\Lambda}_{p,c}^d)= Q_1(\mathbb{\Lambda}_{p,c}^d)=\max_{\rho}
		J(\rho ,\mathbb{\Lambda}_{p,c}^d) = J\left(\tfrac{I^d}{d},\mathbb{\Lambda}_{p,c}^d\right)  
	\nonumber \\ 	&= \log{d}+S\big((1-p)\sigma_0+p\sigma_1\big)-t(p,d^2,\sigma_0,\sigma_1)\;. \label{QVALUE} 
	\end{eqnarray}
	
	\textit{Upper bounds for the DC quantum capacity}.--	
	According to Eq.~(\ref{NICEBOUND}), 
the quantum capacity of the DC $\Lambda_p^d$ can be upper bounded by 
the capacity of~$\mathbb{\Lambda}_{p,c}^d$, irrespectively from the choice we make on 
	the parameter $c$, as long as the degradability  constraint (\ref{deg regime1}) holds true.
	Intuitively however, as $c$ gets larger, the bound gets better, because channel~(\ref{the channel}) gets closer to $\Lambda_p^d$. 
To get the best upper bound for the quantum capacity of $\Lambda_p^d$ we  hence set 	$c=c(p)$.
Accordingly, using the expression for $t(p,d^2,\sigma_0,\sigma_1)$ computed in the SM, our best way to upper bound $Q(\Lambda_p^d)$ is provided by 
	\begin{align}\label{quantum cap}
		&Q(\Lambda_p^d)
		\leq Q(\mathbb{\Lambda}_{p,c(p)}^d)
		=\log{d}+\eta\left(\tfrac{1}{2}\right)\nonumber\\&-\eta\left(\tfrac{1}{2}-\tfrac{(d^2-1)p}{d^2}\right)-(d^2-1)\eta\left(\tfrac{p}{d^2}\right)\: ,
	\end{align} where $\eta(z):=-z\log(z)$
	(as discussed  in the SM, an alternative bound can 
	 be obtained  by choosing the flag states to be pure. The resulting expression is however  much more
	 involved than~(\ref{quantum cap}) and a numerical check reveals that it is less performing 
	 than the latter).

In order to test the quality of our finding we now proceed with a comparison with the limits previously proposed in the literature.
We start considering first the low noise regime ($p\ll1$) where~(\ref{quantum cap})
gives
\begin{equation} 
	Q(\mathbb{\Lambda}_{p,c(p)}^d)=\log{d}+\tfrac{d^2-1}{d^2} \left(\log \left(\tfrac{p}{d^2}\right)-\log{e}+1\right) p+O\left(p^2\right).
\end{equation}
For $d=2$, the above expression is less tight if compared with 
 the bound of Ref.~\cite{lownoiseQ} which for this special regime implies 
\begin{equation} \label{UPl1} 
	Q(\Lambda^2_{p})\leq Q(\mathbb{\Lambda}_{p,c(p)}^d) -\tfrac{3}{4}p+ O\left(p^2\log p\right)\;.
\end{equation}
Things however change when we move out from the $d=2$, low noise
regime. 
To our knowledge, there are two bounds obtained from the degradable extension of the $d$ dimensional depolarizing channel. The first one is 
given in Ref.~\cite{YIN UPP B DEP}  and consists in the following expression

\begin{figure*}[t!]
\includegraphics[width=2\columnwidth]{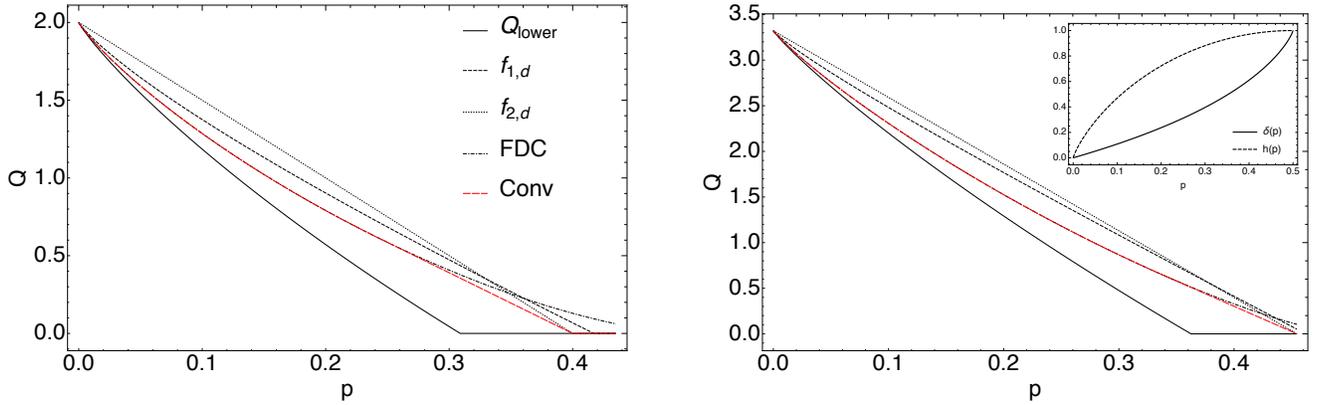} 
\caption{
 $d=4$ (left) and $d=10$ (right): $Q_{lower}$  is the lower bound from Eq.~(\ref{gap1}), $f_{1,d}$ and $f_{2,d}$ are previous bounds from degradable extensions (see Eq.~(\ref{UP1}) and (\ref{UP2})), FDC is our bound presented in Eq.~(\ref{quantum cap}), while finally Conv is the convex hull of all the bounds defined in Eq.~\ref{convhull}. Inset: comparison for the $\order{1}$ gaps for large $d$ between upper bounds and the hashing lower bound, as a function of $p$. For previous bounds the gap is $h(p)$, for ours it is given by the function~(\ref{gap1}). }\label{dlargebounds}
\end{figure*}

\begin{figure}[t!]
	\begin{tabular}{ c }
		\includegraphics[width=1\columnwidth]{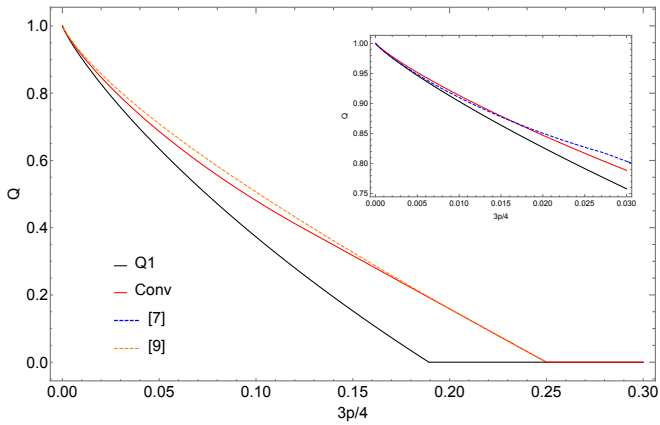} 	\end{tabular}
	\caption{\label{2dbounds}$d=2$ case: $Q_{1}$  is the lower bound from Eq.~(\ref{gap1}), Conv is the convex hull of all the bounds defined in Eq.~\ref{convhull}. In the main plot, we compare our bound with ~\cite{distdepo}, in the inset with ~\cite{SUTT UPP B DEP} at low noise.}
\end{figure}

\begin{align} \label{UP1} 
&Q(\Lambda_p^d)\leq f_{1,d}(p) :=	\eta\Bigl( \tfrac{1+(d-1)\gamma}{d} \Bigr) + (d-1)\eta\Bigl( \tfrac{1-\gamma}{d} \Bigr) \nonumber\\&-\eta\Bigl(1- \tfrac{(d-1) \gamma}{d} \Bigr) 
	- (d-1) \eta\Bigl( \tfrac{ \gamma}{d} \Bigr)\: ,
\end{align}
with $\gamma=\tfrac{2d}{d^2-1}\left(\sqrt{1-p\tfrac{d^2-1}{d^2}}-\left(1-p\tfrac{d^2-1}{2}\right)\right)$.
The second one was instead  
obtained by using the fact that $\Lambda_{p}^d$ is degradable and anti-degradable when $p=\tfrac{d}{2(d+1)}$, see~\cite{YIN UPP B DEP,Dagmar}. Using this fact,~\cite{SS UPP B DEP,YIN UPP B DEP} showed we have 
\begin{equation}  \label{UP2} 
Q(\Lambda_p^d)\leq f_{2,d}(p) :=  \big(1-\tfrac{2p(d+1)}{d}\big) \log{d}\: .
\end{equation}

Given that all of these bounds, including our bound, are obtained from degradable extensions of DCs and the convexity of upper bounds obtained from degradable extensions~\cite{SS UPP B DEP}, we can obtain the following upper bound (see the SM for the detailed proof)
\begin{equation}\label{convhull}
	Q(\Lambda_p^d)\leq \text{conv}\big\{ Q(\mathbb{\Lambda}_{p,c(p)}^d),f_{1,d}(p),f_{2,d}(p) \big\}\: ,
\end{equation}
where the convex hull conv$\{g_1(p),g_2(p),...\}$ is defined as the maximal convex function that is less than or equal to all the $g_i(p)$s.
Figure~\ref{2dbounds} compares the new bound with previous benchmarks for $d=2$, \cite{SUTT UPP B DEP,lownoiseQ} for low noise and \cite{distdepo} for high noise, showing that the new bound is better in an intermediate regime.
Figure~\ref{dlargebounds} represents $Q(\mathbb{\Lambda}_{p,c(p)}^d), f_{1,d}(p), f_{2,d}(p)$, and the convex hull for $d=4$ and $d=10$. 

To be more quantitative, we can study the asymptotic expansion of the capacities of the various extensions for large $d$. Defining $\delta(p):=\eta\left(\tfrac{1}{2}\right)- \eta\left(\tfrac{1}{2}-p\right)+\eta(1-p)$ one can show that
\begin{align} \nonumber 
	Q(\mathbb{\Lambda}_{p,c(p)}^d)&=(1-2p) \log d -h(p)+\delta(p)+\order{\tfrac{1}{\log d}}\;,
\\ f_{1,d}(p)&=(1-2p) \log d+\order{\tfrac{\log d}{d}}\;, \nonumber  
\\ f_{2,d}(p)&=(1-2p) \log d+\order{\tfrac{\log d}{d}}\;, \nonumber  
\end{align}
which should be compared with the lower bound of $Q(\Lambda^d_{p})$ one get by taking the value of the single shot coherent information evaluated on the
completely mixed state, i.e.  
\begin{eqnarray}
	Q({\Lambda}_{p}^d) &\geq& Q_{lower}(\Lambda^d_{p}):= J(\tfrac{I^d}{d}, \Lambda_p^d)  
	\label{lower} 
		\nonumber\\ &=&\log{d}-\eta\left(1-p+\tfrac{p}{d^2}\right)-(d^2-1)\eta\left(\tfrac{p}{d^2}\right)\nonumber\\&=&(1-2p) \log d-h(p)+\order{\tfrac{1}{\log d}}\: .
	\end{eqnarray} 
As we can see, our bound is the only one that shows an $\order{1}$ term which is not zero (and negative). Furthermore, the gap between our bound and the lower bound scales as
\begin{align} \label{gap1} 
Q(\mathbb{\Lambda}_{p,c(p)}^d)-Q_{lower} (\Lambda^d_{p})=\delta(p) +\order{\tfrac{1}{\log d}}\;.
\end{align}
On the contrary the differences between the other upper bounds and the lower bound exhibit
a $\order{1}$ gap equal  to $h(p)$ which, as shown in Fig.~\ref{dlargebounds} is larger than (\ref{gap1}) 
for $p<\tfrac 1 2$ (where the quantum capacity is not zero). In particular, it appears that our
inequality gives a much better bound for low $p$, since $h(p)$ has derivative that diverges as $-\log p$ when $p\rightarrow 0$, while $\delta (p)$ scales 
linearly in $p$. 

	\textit{Discussion}.--We introduced a specific flagged version of DC which  for a certain values of the parameter is degradable
	allowing us to compute analytic bound for the quantum capacity of the original map.
 Our result works in any dimension, and it is the tightest available analytical upper bound. Unlike other degradable extensions of depolarizing channel~\cite{SS UPP B DEP,YIN UPP B DEP}, the introduced flags are not orthogonal. 
	The idea we used is of general applicability and could give new good bounds for many other channels.
	
	\textit{Acknowledgement}.--We thank Felix Leditzky, Andreas Winter, and Mark Wilde for helpful feedbacks.

\newpage

\appendix
\begin{widetext}

\section{Explicit value of $t(p,d,\sigma_0,\sigma_1)$ }
The fact that $\mathbb{\Lambda}_{p}^d$ is covariant under $SU(d)$ implies that
the output von Neumann entropy associated with a generic input state is a constant 
$t(p,d,\sigma_0,\sigma_1)$ 
that explicitly does not depend upon the specific value of $|\psi\rangle$ but only upon
the parameters that characterize the map, i.e. $p$, $\sigma_0$, $\sigma_1$ and $d$.
A simple algebra  permits us to explicit determine the value of $t(p,d,\sigma_0,\sigma_1)$
obtaining 	\begin{eqnarray} \label{deft} 
t(p,d,\sigma_0,\sigma_1) &:=& S(\mathbb{\Lambda}_{p}^d[\ket{\psi}\bra{\psi}])=S\left((1-p)\ket{\psi}\bra{\psi}\otimes\sigma_0+p\tfrac{I^d}{d}\otimes\sigma_1\right)\\ \nonumber
&=&
h\left(\tfrac{d(1-p)+p}{d}\right) +\tfrac{p(d-1)}{d}\log(d-1) + \tfrac{d(1-p)+p}{d} S\left(\tfrac{(1-p)\sigma_0+\tfrac{p}{d}\sigma_1}{\tfrac{d(1-p)+p}{d}}\right)+ \tfrac{p(d-1)}{d}S\left(\sigma_1\right), 
\end{eqnarray} 
where $h(x):=-x \log x -(1-x)\log(1-x)$ is the binary entropy. 
	\section{Classical capacities of the FDC}
	By convexity of 
	the von Neumann entropy, it follows that   
\begin{eqnarray} \label{SMIN} 
\min_\rho S(\mathbb{\Lambda}_{p}^d[\rho]) = t(p,d,\sigma_0,\sigma_1) \;.\end{eqnarray}
Using above observation we compute  the Holevo capacity of the map $C_1(\mathbb{\Lambda}_{p}^d)$. Notice that for any ensemble $\{p_i;\rho_i\}$, one can create a larger ensemble $\{p_i, dU; U\rho_i U^\dagger\}$, where the state $U\rho_iU^\dagger$ is extracted with probability density $p_idU$,
	 where $dU$ is the Haar measure of $SU(d)$. 
	By the concavity of the Holevo quantity it follows 
		\begin{align}\label{concavity_covariant}
	&\chi(\{p_i;\mathbb{\Lambda}_{p}^d[\rho_i]\})\leq \chi(\{p_i,dU; \mathbb{\Lambda}_{p}^d[U\rho_iU^\dag]
	\})
	=\log d +S((1-p)\sigma_0+p\sigma_1)-\sum_{i}p_iS(\mathbb{\Lambda}_{p}^d[\rho_i])\;, 
	\end{align}
where we used the depolarizing identity $\int dU \;U \rho U^\dagger=\tfrac{I^d}{d}.$

We can now invoke (\ref{SMIN}) to put an upper bound on $\chi(\{p_i,dU; \mathbb{\Lambda}_{p}^d[U\rho_iU^\dag]
	\})$ by replacing all the $S(\mathbb{\Lambda}_{p}^d[\rho_i])$ terms with the constant 
	$t(p,d,\sigma_0,\sigma_1)$. The resulting quantity no longer depends on the input of the channel and provide an achievable maximum for the Holevo information of the channel yielding the identity 
	\begin{equation}
	C_1(\mathbb{\Lambda}_{p}^d)=\log d + S((1-p)\sigma_0+p\sigma_1)-t(p,d,\sigma_0,\sigma_1),
	\end{equation}
(the achievability being granted e.g. by  ensembles of the form $\{ dU; U|\psi\rangle\langle \psi| U^\dagger\}$, with $|\psi\rangle$ arbitrarily chosen).

To compute the entanglement assisted capacity of $\mathbb{\Lambda}_{p}^d$, we use the fact that the quantum mutual information of a channel 
 is concave in $\rho$ \cite{HOL BOOK}.
Exploiting this and the covariance of $\mathbb{\Lambda}_{p}^d$ under $SU(d)$ we can then write 
	\begin{align}
	&I\left(\tfrac{I^d}{d},\mathbb{\Lambda}_{p}^d\right) =	I\left(\int U\rho U^\dagger\: dU,\mathbb{\Lambda}_{p}^d\right)\geq \int I\left(U\rho U^\dagger,\mathbb{\Lambda}_{p}^d\right) \: dU= I(\rho,\mathbb{\Lambda}_{p}^d)\: . \label{conc11} 
	\end{align}
	Therefore, we can conclude that  the state that maximizes the quantum mutual information is $\tfrac{I}{d}$ 
	and after some algebra we get
	\begin{eqnarray}
		C_E(\mathbb{\Lambda}_{p}^d)&=&I\left(\tfrac{I^d}{d},\mathbb{\Lambda}_{p}^d\right)=2\log{d}+S\big((1-p)\sigma_0+p\sigma_1\big) 		-t(p,d^2,\sigma_0,\sigma_1), 
\label{NEW} 
	\end{eqnarray}

\section{Detailed analysis of degradability}
	To find complementary channel $\tilde{\mathbb{\Lambda}}_{p,c}^d$ we should first write the Stinespring~\cite{HOL BOOK,WILDE BOOK}  form of $\mathbb{\Lambda}_{p,c}^d$ as it is discussed in Eq.~(\ref{steinspring}). For this purpose we add extra degree of freedom
extending the environment
Hilbert space to $\mathcal{H}_{E}=\mathcal{H}_{1}\otimes\mathcal{H}_2\otimes\mathcal{H}_{3}\otimes\mathcal{H}_{4}\otimes\mathcal{H}_{5}$ where  $\mathcal{H}_{1}$, $\mathcal{H}_{2}$ and $\mathcal{H}_{5}$ are two dimensional, and $\mathcal{H}_{3}$ and $\mathcal{H}_{4}$ are $d$ dimensional Hilbert spaces. Simple algebra can hence be used to 
verify that the Stinespring representation of the channel can be obtained through the following unitary interaction  	\begin{equation}
U_{AE}\ket{\psi}_A\ket{0}_1\ket{0}_2\ket{\Phi^d}_{3,4}\ket{0}_5=\sqrt{1-p}\ket{\psi}_A\vert\sigma_0\rangle\!\rangle_{1,2} \ket{\Phi^d}_{3,4}\ket{0}_5+\sqrt{p}\ket{\Phi^d}_{A,4}\ket{e_1}_1\ket{e_1}_2\ket{\psi}_3\ket{1}_5 \: , \label{app unitary} 
\end{equation}
where $\ket{0},\ket{1}$ are two orthogonal states, $\ket{\Phi^d}$   is a maximally entangled state in dimension $d$, and $\vert\sigma_0\rangle\!\rangle_{1,2}$ is a purifications of $\sigma_0$, 
and the trace in Eq.~(\ref{steinspring})  is on labels 2,3,4,5 (see the next section in the SM for the details).  To find the complementary channel instead of taking trace over states 2,3,4,5 we should take trace over states A,1. Carrying out the calculation we get
\begin{align} \nonumber 
\tilde{\mathbb{\Lambda}}_{p,c}^d(\ket{\psi}\bra{\psi})=&(1-p){\sigma_0}_2\otimes\ket{\Phi^d}\bra{\Phi^d}_{3,4}\otimes\ket{0}\bra{0}_5+p{\ket{e_1}\bra{e_1}}_2\otimes\ket{\psi}\bra{\psi}_3\otimes\tfrac{I^d_4}{d}\otimes\ket{1}\bra{1}_5\\
&+\sqrt{p(1-p)}\big[c \,\tr_A(\ket{\psi}_A\ket{\Phi^d}_{3,4}\ket{0}_5\bra{\Phi^d}_{A,4}\bra{\psi}_3\bra{1}_5)+h.c.\big]\otimes\ket{e_1}\bra{e_1}_2 \: . \label{def compl1} 
\end{align}
We now look for the existence of a degrading CPTP channel  $W_{p,c}$ connecting $\mathbb{\Lambda}_{p,c}^d$ and $\tilde{\mathbb{\Lambda}}_{p,c}^d$, i.e.  satisfying the condition
$W_{p,c}\circ \mathbb{\Lambda}_{p,c}^d=\tilde{\mathbb{\Lambda}}_{p,c}^d$ or explicitly 
\begin{equation}\label{condition w}
(1-p)W_{p,c}(\rho\otimes{\sigma_0}_1)+\tfrac{p}{d}W_{p,c}(I^d\otimes\ket{e_1}\bra{e_1}_1)=\tilde{\mathbb{\Lambda}}_{p,c}^d(\rho)\: .
\end{equation}
As a suitable candidate for  $W_{p,c}$ we consider  a two-step process which first performs a measurement on system 1 that then triggers an action on $A$. Specifically for the measurement we assume an orthogonal projection in the basis $\ket{e_1}$ and $\ket{e^\perp_1}$. For the action on $A$ instead
we assume that  if the measurement outcome is  $\ket{e_1}$ we will prepare whatever state was left on $A$ into the fixed state ${{\ket{e^\perp_1}\bra{e^\perp_1}}_2}\otimes\ket{\Phi^d}\bra{\Phi^d}_{3,4}\otimes\ket{0}\bra{0}_5$; on the contrary, if the result is $\ket{e^\perp_1}$ we operate on  $A$ with a channel of the form 
$\tilde{\mathbb{\Lambda}}_{q,c'}^d$ 
with properly selected  parameters $q,c'$. With this choice, the resulting mapping  $W_{p,c}$ on $\rho_{A,1}$ is  hence given by 
\begin{equation}\label{def w}
W_{p,c}(\rho_{A,1}):=\bra{e_1}\tr_{A}(\rho_{A,1})\ket{e_1}{{\ket{e^\perp_1}\bra{e^\perp_1}}_2}\otimes\ket{\Phi^d}\bra{\Phi^d}_{3,4}\otimes\ket{0}\bra{0}_5+\bra{e^\perp_1}\tr_{A}(\rho_{A,1})\ket{e^\perp_1}\tilde{\mathbb{\Lambda}}_{q,c'}^d(\tr_{1}(\rho_{A,1}))\: .
\end{equation}
With this choice the  condition~(\ref{condition w}) becomes  
\begin{align}
(1-p)\big[c^2{{\ket{e^\perp_1}\bra{e^\perp_1}}_2}\otimes\ket{\Phi^d}\bra{\Phi^d}_{3,4}\otimes\ket{0}\bra{0}_5+(1-c^2)\tilde{\mathbb{\Lambda}}_{q,c'}^d(\rho_A)\big]+p{{\ket{e^\perp_1}\bra{e^\perp_1}}_2}\otimes\ket{\Phi^d}\bra{\Phi^d}_{3,4}\otimes\ket{0}\bra{0}_5=\tilde{\mathbb{\Lambda}}_{p,c}^d(\rho_A)\: .
\end{align}
which can be satisfied if it is possible to find $q,c'
\in [0,1]$  such that
\begin{align}
q=\tfrac{p}{(1-p)(1-c^2)}\:, \qquad \quad {c'}^2=\tfrac{c^2(1-p)}{1-2p-c^2+pc^2}\: .
\end{align}
Doing simple algebra reveals that this is the case  for all those cases where the following inequality holds, 
\begin{equation}\label{deg regime}
c \leq\sqrt{\tfrac{1-2p}{2-2p}} \: .
\end{equation}
Under this condition the  channel $\mathbb{\Lambda}_{p,c}^d$ is degradable
	
\section{Stinespring representation  and 
complementary channel}\label{stinespring proof}
Here we show that the mapping
	\begin{equation}\label{app unitary}
U_{AE}\ket{\psi}_A\ket{0}_1\ket{0}_2\ket{\Phi^d}_{3,4}\ket{0}_5=\sqrt{1-p}\ket{\psi}_A\vert\sigma_0\rangle\!\rangle_{1,2} \ket{\Phi^d}_{3,4}\ket{0}_5+\sqrt{p}\ket{\Phi^d}_{A,4}\vert\sigma_1\rangle\!\rangle_{1,2}\ket{\psi}_3\ket{1}_5 \: ,
\end{equation}
provides a Stinespring representation of the channel $\Lambda^d_{p,\sigma_0,\sigma_1}$.
For this purpose we first notice that (\ref{app unitary}) identifies a unitary transformation because  in the domain where we have defined
it does preserve the scalar product: indeed introducing the compact notation 
 $\ket{\psi,e}_{AE}:=\ket{\psi}_A\ket{0}_1\ket{0}_2\ket{\Phi^d}_{3,4}\ket{0}_5$ we have 
\begin{equation}
	{_{AE}\bra{\phi,e}}U^\dagger_{AE}U_{AE}\ket{\psi,e}_{AE}=(1-p)\;  {_A\langle}{\phi}
	|{\psi}\rangle_A
	+p\;  {_3\langle}{\phi}
	|{\psi}\rangle_3  ={_A\langle}{\phi}
	|{\psi}\rangle_A = {_{AE}\langle}{\phi,e}
	|{\psi,e}\rangle_{AE}  \: .
\end{equation}
Next we notice that by tracing over $2,3,4,5$ we get 
\begin{eqnarray}
\tr^{(A1)}  \left[ U_{AE}\ket{\psi,e}_{AE}\bra{\psi,e}U^\dagger_{AE}\right] = (1-p) \ket{\psi}_{A}\bra{\psi} \otimes\sigma_0+p\tfrac{I^d}{d}\otimes\sigma_1
= \Lambda^d_{p,\sigma_0,\sigma_1}( \ket{\psi}_{A}\bra{\psi} )\;,
\end{eqnarray} 
for all possible input state $\ket{\psi}_A$ (here $\tr^{(A1)}$ indicates that we are taking the partial trace with respect to all degree of freedom of the system but $A,1$). 

From the above definition we now show that Eq.~(\ref{def compl1}) is the complementary 
channel~(\ref{def compl}) of $\mathbb{\Lambda}_{p,c}^d$, i.e. that the following identity holds 
true 
\begin{equation}
	\tilde{\mathbb{\Lambda}}_{p,c}^d
	[\rho]=\tr_{A1}\Big[ U_{AE}(\rho\otimes\ket{e}\bra{e}_E) U_{AE}^\dagger\Big]\: ,
\end{equation}
for all input states $\rho$ (here $\tr_{A1}$ indicates that the trace is taken on $A$ and $1$). 
Without loss of generality we can always focus on pure input states. Under this condition
 the right side of the previous expression yields
\begin{align}
	U_{AE} \left( | \psi\rangle_A\langle \psi| \otimes\ket{e}\bra{e}_E \right) U_{AE}^\dagger
	=&(1-p)\ket{\psi}\bra{\psi}_A\otimes\vert\sigma_0\rangle\!\rangle\langle\!\langle\sigma_0\vert_{1,2}\otimes \ket{\Phi^d}\bra{\Phi^d}_{3,4}\otimes\ket{0}\bra{0}_5\\
	&+ p\ket{\Phi^d}\bra{\Phi^d}_{A,4}\otimes\ket{e_1}\bra{e_1}_1\otimes\ket{e_1}\bra{e_1}_2\otimes\ket{\psi}\bra{\psi}_3\otimes\ket{1}\bra{1}_5	\\
	&+ \sqrt{p(1-p)}\ket{\psi}_A \ket{\Phi^d}_{3,4}\bra{\Phi^d}_{A,4}\bra{\psi}_3\otimes\vert\sigma_0\rangle\!\rangle_{1,2}\bra{e_1}_1\bra{e_1}_2\otimes\ket{0}\bra{0}_5+h.c. \: 
\end{align}
Given that $\vert\sigma_0\rangle\!\rangle_{1,2}=c\ket{e_1}\ket{e_1}_{1,2}+\sqrt{1-c^2}\ket{e^\perp_1}\ket{e^\perp_1}_{1,2}$ we can take trace over $2,3,4,5$ and get
	\begin{align}
\tilde{\mathbb{\Lambda}}_{p,c}^d(\ket{\psi}_A\bra{\psi})=&(1-p){\sigma_0}_2\otimes\ket{\Phi^d}\bra{\Phi^d}_{3,4}\otimes\ket{0}\bra{0}_5+p{\ket{e_1}\bra{e_1}}_2\otimes\ket{\psi}\bra{\psi}_3\otimes\tfrac{I^d_4}{2}\otimes\ket{1}\bra{1}_5\\
&+\sqrt{p(1-p)}\big[c \,\tr_A(\ket{\psi}_A\ket{\Phi^d}_{3,4}\ket{0}_5\bra{\Phi^d}_{A,4}\bra{\psi}_3\bra{1}_5)+h.c.\big]\otimes\ket{e_1}\bra{e_1}_2 \: .
\end{align}

\section{Pure  flags expansion}\label{pure}
The condition for degradability for the pure flags are  similar to the case where the flags are mixed but diagonal. In this scenario the channel explicitly writes as 
	\begin{equation}\label{the channel 2}
{\mathbb{\Lambda}}'^d_{p,c}[\cdots]=(1-p)[\cdots] \otimes\ket{e_0}\bra{e_0}+p\tfrac{I^d}{d}\otimes\ket{e_1}\bra{e_1}\: ,
\end{equation}
where the parameter $c$ refers now to the overlap $c:=\bra{e_1}\ket{e_0}$. Notice that the phase in $c$ is not important in studying the degradability of $\Lambda'^d_{p,c}$ since the phase in $c$ can be set to zero by acting with a
 unitary transformation after the action of the channel~(\ref{the channel 2}): accordingly in the following 
 we shall assume  $c$ to be real without loss of generality. 
 
 To find complementary channel ${\mathbb{\Lambda}}'^d_{p,c}$ we should first write the Stinespring form of this transformation. The Hilbert space of the environment is decomposed as $\mathcal{H}_{E}=\mathcal{H}_{1}\otimes\mathcal{H}_{2}\otimes\mathcal{H}_{3}\otimes\mathcal{H}_{4}$ where $\mathcal{H}_{1}$ and $\mathcal{H}_{4}$ are two dimensional, and $\mathcal{H}_{2}$ and $\mathcal{H}_{3}$ are $d$ dimensional Hilbert spaces. The unitary interaction between system and environment acts as following 
\begin{equation}
U'_{AE}\ket{\psi}_A\ket{0}_1\ket{\Phi^d}_{2,3}\ket{0}_4=\sqrt{1-p}\ket{\psi}_A\ket{e_0}_1\ket{\Phi^d}_{2,3}\ket{0}_4+\sqrt{p}\ket{\Phi^d}_{A,3}\ket{e_1}_1\ket{\psi}_2\ket{1}_4 \: ,
\end{equation}
where $\ket{0},\ket{1}$ are two orthogonal states, $\ket{\Phi^d}$   is a maximally entangled states in dimension $d$, $\ket{e}_E=\ket{0}_1\ket{\Phi^d}_{2,3}\ket{0}_4$, and the trace in Eq.~(\ref{steinspring}) here is on states 2,3,4. Doing simple calculation we can show that this is a Stinespring representation of \ref{the channel 2}. To find the complementary channel instead of taking trace over states 2,3,4 we should take trace over states A,1, carrying out the calculation we get
\begin{align}
\tilde{\mathbb{\Lambda}}'^d_{p,c}
(\ket{\psi}\bra{\psi})=&(1-p)\ket{\Phi^d}\bra{\Phi^d}_{2,3}\otimes\ket{0}\bra{0}_4+p\ket{\psi}\bra{\psi}_2\otimes\tfrac{I^d_3}{2}\otimes\ket{1}\bra{1}_4\\
&+\sqrt{p(1-p)}\big[c \,\tr_A(\ket{\psi}_A\ket{\Phi^d}_{2,3}\ket{0}_4\bra{\Phi^d}_{A,3}\bra{\psi}_2\bra{1}_4)+h.c.\big] \: .
\end{align}
As the form of $\tilde{\mathbb{\Lambda}}'^d_{p,c}$ is exactly the same as $\tilde{\mathbb{\Lambda}}^d_{p,c}$, the regime where ${\mathbb{\Lambda}}'^d_{p,c}$
 is degradable is the same as before, i.e. 
\begin{equation}
	c^2\leq\tfrac{1-2p}{2-2p} \: .
\end{equation}
In this regime 
 the quantum capacity of ${\mathbb{\Lambda}}'^d_{p,c}$
  can be computed as in Eq.~(\ref{QVALUE}), i.e. 
	\begin{equation}
Q({\mathbb{\Lambda}}'^d_{p,c})=\log{d}+S\big((1-p)\ket{e_0}\bra{e_0}+p\ket{e_1}\bra{e_1}\big)-t(p,d^2,\ket{e_0},\ket{e_1})\: ,
\end{equation}
which after some algebra can be casted into the expression 
	\begin{align} 
Q({\mathbb{\Lambda}}'^d_{p,c})&=\log{d}+\eta[\tfrac{1}{2} \left(1-\sqrt{-2 (p-1) p \cos (\theta)+2 (p-1) p+1}\right)]+\eta[\tfrac{1}{2} \left(1+\sqrt{-2 (p-1) p \cos (\theta)+2 (p-1) p+1}\right)]\nonumber \\ 
&-\eta[\tfrac{d^2 (-p)+d^2-\sqrt{d^4 p^2-2 d^4 p+d^4-2 d^2 p^2 \cos (\theta )+2 d^2 p \cos (\theta )+p^2}+p}{2 d^2}] \nonumber \\
&-\eta[\tfrac{d^2 (-p)+d^2+\sqrt{d^4 p^2-2 d^4 p+d^4-2 d^2 p^2 \cos (\theta )+2 d^2 p \cos (\theta )+p^2}+p}{2 d^2}]+\tfrac{p(d^2-1)}{d^2}\log(\tfrac{p}{d^2})\:, \label{NEWBOUND} 
\end{align}
where $\cos(\theta)=2c^2-1$ and $\eta(z):=-z\log(z)$.

\section{Combination of different bounds from degradable extensions}

In this section we present one of the results in Ref.~\cite{SS UPP B DEP}.
We call $N$ a degradable extension of $\Lambda$ if $N$ is degradable and there is a second channel $R$ such that $R\circ N=\Lambda$. In  Ref.~\cite{SS UPP B DEP} it has been shown that if $N_0$ is a degradable extension of $\Lambda_0$ and $N_1$ is a degradable extension of $\Lambda_1$ then $N=\lambda N_0\otimes\ket{0}\bra{0}+(1-\lambda)N_1\otimes \ket{1}\bra{1}$
is a degradable extension of $\Lambda=\lambda\Lambda_0+(1-\lambda)\Lambda_1$ for every $0\leq \lambda \leq 1$, and the quantum capacities satisfy the following relation
\begin{align}\label{conv}
Q(\Lambda)\leq Q_1(N)\leq \lambda Q_1(N_0)+(1-\lambda) Q_1(N_1)\;. 
\end{align}
This theorem can be used to show if we have upper bounds for the quantum capacity
of two channels, all obtained from degradable extensions,
the convex combination of the bounds is also an upper bound for the respective convex combination of the channels. We clarify this with an example: Consider the depolarizing channel i.e.  $	\Lambda_p^d[\cdots] = (1-p)[\cdots]+p\Tr[\cdots]\tfrac{I^d}{d}$. The set of all values of $p$ for which $\Lambda_{p}^d$ is a CPTP is $P$, and $N_p$ is a degradable extension of $\Lambda_{p}^d$ for all $p\in P$. If $p_0,p_1 \in P$, then  
$N_{p_0},N_{p_1}$ are degradable extensions of $\Lambda_{p_0}^d, \Lambda_{p_1}^d$ respectively, then
\begin{equation}
Q(\Lambda_{p_0}^d)\leq g(p_0):=Q_1(N_{p_0}),\quad Q(\Lambda_{p_1}^d)\leq g(p_1):=Q_1(N_{p_1})\, .
\end{equation}
Therefore
\begin{equation}
N=\lambda N_{p_0}\otimes\ket{0}\bra{0}+(1-\lambda)N_{p_1}\otimes \ket{1}\bra{1}\: ,
\end{equation}
is a degradable extension of $\Lambda^d_{\lambda p_0+(1-\lambda)p_1}$, then using~\ref{conv} we get $Q(\Lambda^d_{\lambda p_0+(1-\lambda)p_1})\leq \lambda g(p_0)+(1-\lambda) g(p_1).$
As this holds for all $p_0, p_1\in P$, therefore conv\{$g(p)$\} is also an upper bound for the quantum capacity of $\Lambda_p^d$, where
\begin{align*}
\text{conv}\{g(p)\}&:=\inf\limits_{\substack{p_0,p_1\in P,\\ 0\leq\lambda\leq 1}}\{ \lambda g(p_0)+(1-\lambda)g(p_1):p=\lambda p_0+(1-\lambda)p_1 \}\: .
\end{align*}
{In particular}, given $g_1(p),...,g_n(p)$, all upper bounds for the quantum capacity of depolarizing channel all derived from degradable extensions, then $g_{\text{min}}(p):=\min\{g_1(p),...,g_n(p)\}$ is also an upper bound and therefore
$	\text{conv}\{g_1(p),...,g_n(p)\}:=\text{conv}\{g_\text{min}(p)\}, $
is also an upper bound too.

\end{widetext}
\end{document}